\documentclass[shortnote,twocolumn]{jpsj3}
\usepackage{txfonts}
\usepackage{bm}
\usepackage{braket}

\usepackage{color}

\title{Bosonic Nevanlinna Analytic Continuation}

\author{Kosuke Nogaki$^1$\thanks{nogaki.kosuke.83v@st.kyoto-u.ac.jp} and Hiroshi Shinaoka$^{2,3}$}
\inst{$^1$Department of Physics, Kyoto University, Kyoto 606-8502, Japan \\
$^2$ Department of Physics, Saitama University, Saitama 338-8570, Japan \\
$^3$ JST, PRESTO, 4-1-8 Honcho, Kawaguchi, Saitama 332-0012, Japan}

\abst{
Analytical continuation (AC) connects theoretical calculations and experimentally measurable quantities. The recently proposed Nevanlinna AC method is capable of accurately reproducing the sharp features of spectral functions at high frequencies while maintaining the causality of the response function. However, their use is currently limited to fermions. Here, we present an extension of this method to bosons using the hyperbolic tangent trick, allowing us to transform bosons into auxiliary fermions to which the Nevanlinna analytic continuation can be applied.
}

\begin{document}
\maketitle

In the field of finite-temperature quantum field theories, many sophisticated numerical techniques such as a variety of perturbation theories~\cite{AGD}, lattice-~\cite{Blankenbecler1981}, continuous-time~\cite{Gull2011} quantum Monte Carlo methods, and lattice quantum chromodynamics~\cite{Gattringer} have been developed. 
Because they are formulated in imaginary time, their outputs are typically a Matsubara correlation function $\mathcal{G}(i\omega_n)$.
To extract information on experimentally observable quantities, analytic continuation (AC) from the Matsubara correlation function $\mathcal{G}(i\omega_n)$ to the retarded correlation function $G^{\mathrm{R}}(\omega)$ is required.
The spectral function $\rho(\omega) = - (1/\pi) \mathrm{Im}G^{\mathrm{R}}(\omega)$ describes the system dynamics.

Because numerical AC is known to be an ill-conditioned problem, directly solving the inverse problem is bound to fail.
To overcome this difficulty, many efforts have been made to develop approximate methods, such as the Pad\'{e} approximation~\cite{Baker}, maximum entropy method (MaxEnt)~\cite{Bryan1990,Jarrell1996}, stochastic AC~\cite{Sandvik1998,Mishchenko2000,Vafayi2007,Fuchs2010,Goulko2017}, machine learning approaches~\cite{Yoon2018}, genetic algorithms~\cite{Vitali2010}, sparse modeling methods (SpM)~\cite{Otsuki2017,Otsuki2020}, and the pole-fitting approach~\cite{Huang2022}.
However, an empirical interpolation, such as the Pad\'{e} approximation, sometimes breaks causality, and a fitting approach, such as MaxEnt, cannot capture the sharp features of the spectral function in a large-$\omega$ region.

Recently, Fei \textit{et al.}~\cite{Fei2021,Fei2021_2} proposed a mathematically rigorous AC method based on Nevanlinna theory.
In this method, the analytical structure of fermionic correlation functions plays an essential role. Namely, the Matsubara and retarded correlation functions are classified into the class of Nevanlinna functions that map the open upper-half plane to the closed upper-half plane.
The necessary and sufficient condition for the existence of Nevanlinna interpolants is known as the positive semidefiniteness of the Pick matrix constructed from the input data~\cite{Pick1917,Khargonekar1985}.
This condition enables the selection of a subset of input data that preserves causality.
Moreover, the Nevanlinna interpolation procedure can be performed efficiently using the Schur algorithm~\cite{Schur1918,Adamyan2003} and can resolve sharp features in a large-$\omega$ region.
Schur algorithm iteratively interpolates the input data, leaving one function $\theta_{M+1}(z)$ undetermined.

To date, the Nevanlinna AC method has been limited to fermionic cases.
However, bosonic correlation functions describe most correlation functions that can be measured in the experiment.
Therefore, the development of the Nevanlinna AC method for bosons is desirable.

In this Short Note, we introduce a trick that converts the bosonic AC problem into an auxiliary fermionic AC problem.
A similar technique has been used with MaxEnt~\cite{Meyer2007} and SpM~\cite{Itou2020} but not with Nevanlinna AC.
We then apply the Nevanlinna AC method to an auxiliary fermionic AC problem and reconstruct a bosonic spectral function by multiplying the auxiliary fermionic spectral function by the hyperbolic tangent. 

\begin{figure}
\includegraphics[keepaspectratio, scale=0.23]{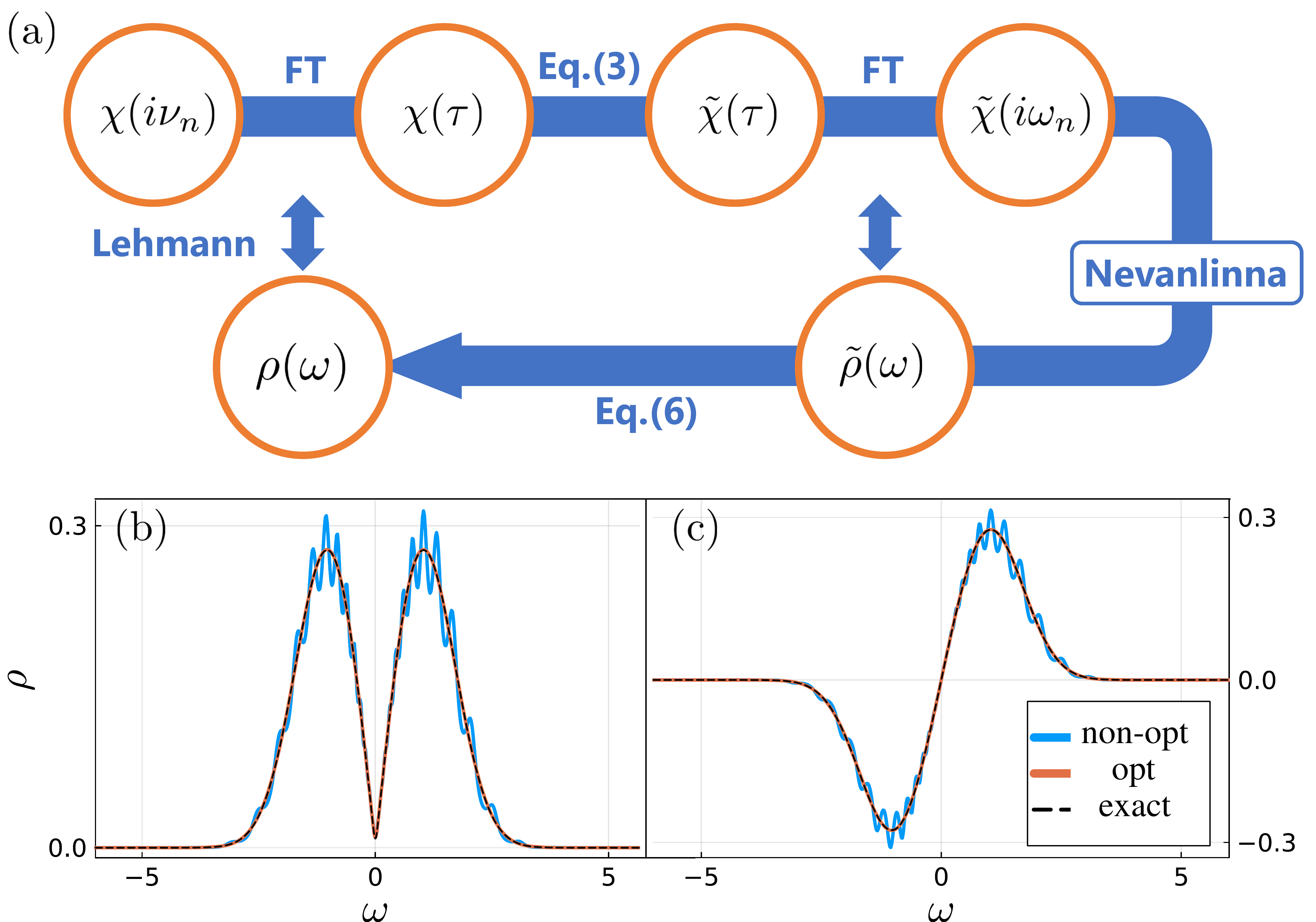}
\caption{(Color online) (a) Procedure for solving a bosonic AC problem from input $\chi(i\nu_n)$.
Here, FT represents the Fourier transform between the imaginary-time and the Matsubara-frequency domains.
The input data $\chi(i\nu_n)$ can be either on a sparse Matsubara grid or a dense one.
(b-c) Exact and reconstructed spectral functions with and without the Hardy optimization. 
The result of a symmetric double-peak model is shown.
We use $\beta= 100$ and the intermediate representation grid with 36 Matsubara positive frequency points~\cite{Shinaoka2017,Li2020,shinaokalecturenote}.
The auxiliary fermionic and bosonic spectral functions are shown in (b) and (c) respectively.}
\label{fig:merge}
\end{figure}

Let us consider the correlation functions between the bosonic operator $A$ and its conjugate $A^\dagger$.
In typical cases, bosonic operators $A$ are even-degree polynomials of creation and annihilation operators for fermions, $c^\dagger_{\mathrm{F}}$, $c_{\mathrm{F}}$ or arbitrary degree polynomials of creation and annihilation operators for bosons, $c^\dagger_{\mathrm{B}}$, $c_{\mathrm{B}}$. 
The Matsubara correlation function of $A$ is defined as
$\chi_{A}(\tau) = -\langle T_{\tau}A(\tau)A^\dagger(0) \rangle$,
where $T_{\tau}$ is the time-ordering operator and $A(\tau)=e^{(H-\mu N)\tau} A e^{-(H-\mu N)\tau}$ with the Hamiltonian $H$, particle number operator $N$, and chemical potential $\mu$.
Here, $\langle \cdots \rangle$ represents the thermal average, namely, $\langle \cdots \rangle := \mathrm{tr}\{e^{-\beta(H-\mu N)}\cdots\}/\Xi$ with inverse temperature $\beta = 1/T$ and partition function $\Xi = \mathrm{tr}\{e^{-\beta(H-\mu N)}\}$.
Owing to the cyclic property of the trace, periodicity $\chi_A(\tau+\beta)=\chi_A(\tau)$ is satisfied for bosonic correlation functions.
Fourier-transformed correlation function
$\chi_{A}(i\nu_n) = \int^\beta_0 \, d\tau \, e^{i\nu_n \tau} \chi_{A}(\tau)$ 
is a central object in finite-temperature quantum field theory, where $i\nu_n=2n\pi iT$ are bosonic Matsubara frequencies.
These correlation functions are related to the spectral function $\rho_{A}(\omega)$ through the Lehmann (spectral) representation
\begin{align}
\label{eq:boson_lehman}
    \chi_A(\tau) &= -\int^\infty_{-\infty} d\omega \, \frac{e^{-\tau\omega}}{1 - e^{-\beta\omega}} \, \rho_{A}(\omega), \\
    \chi_A(i\nu_n) &= \int^\infty_{-\infty} d\omega \, \frac{\rho_{A}(\omega)}{i\nu_n-\omega},
\end{align}
respectively.
Spectral function $\rho_{A}(\omega)$ contains information on the dynamics of the system  probed or measured by operator $A$. 
For example, a one-body correlation function with $A=c \ \mathrm{or} \ c^\dagger$ contains information about the one-particle excitation spectrum.
Similarly, a two-body correlation function with $A=c^\dagger c \ \mathrm{or} \ cc$ contains information about the collective and individual excitation spectra.
The central issue in numerical AC is the estimation $\rho_A(\omega)$ from $\chi_A(i\nu_n)$.

We define an artificial antiperiodic function $\tilde{\chi}(\tau)$,
\begin{align}
    \tilde{\chi}_A(\tau) = \left\{
                        \begin{array}{ll}
                        \chi_A(\tau) & (0 < \tau < \beta)\\
                        -\chi_A(\tau+\beta) & (-\beta < \tau < 0)
                        \end{array}
                        \right. .
    \label{eq:fermion_chi}
\end{align}
Substituting Eq.~(\ref{eq:boson_lehman}) into Eq. ~(\ref{eq:fermion_chi}), the Fourier transform $\tilde{\chi}_A(i\omega_n) = \int^\beta_0 \, d\tau \, e^{i\omega_n \tau} \tilde{\chi}_{A}(\tau)$ is given by:
\begin{align}
    \tilde{\chi}_{A}(i\omega_n) &= -\int^\infty_{-\infty} d\omega \, \int^\beta_0 \, d\tau \, e^{i\omega_n \tau} \frac{e^{-\tau\omega}}{1 + e^{-\beta\omega}} \, \tilde{\rho}_{A}(\omega), \\
                                &= \int^\infty_{-\infty} d\omega \, \frac{\tilde{\rho}_{A}(\omega)}{i\omega_n-\omega},
\end{align}
where $i\omega_n=(2n+1)\pi iT$ are the fermionic Matsubara frequencies~\cite{Meyer2007,Itou2020}.
Here, $\rho_A$ and $\tilde{\rho}_A$ are related as follows: 
\begin{align}
    \rho_A(\omega) = \tilde{\rho}_A(\omega)\cdot\tanh(\beta\omega/2).
\end{align}
and the sum rule of $\tilde{\rho}_A$ is given by  $\int^\infty_{-\infty} \, d\omega \, \tilde{\rho}_A(\omega) = \tilde{\chi}_A(0^+) + \tilde{\chi}_A(\beta-0^+) = S$.
This procedure corresponds to {\it fermionization} of $\chi_A(i\nu_n)$: a bosonic AC with $\rho_A$ is mapped to an auxiliary fermionic AC with $\tilde{\rho}_A$.
Using this technique, we can estimate $\tilde{\rho}_A$ from $\tilde{\chi}_{A}(i\omega_n)$ efficiently using the Nevanlinna AC method~\cite{Fei2021}.  
Figure~\ref{fig:merge}(a) illustrates the procedure.

First, we numerically transform $\chi(i\nu_n)$ into $\tilde{\chi}(i\omega_n)$ through an intermediate representation~\cite{Shinaoka2017,shinaokalecturenote} using sparse sampling. ~\cite{Li2020}
We then estimate $\tilde{\rho}$ from $\tilde{\chi}(i\omega_n)$ using the Nevanlinna AC method. ~\cite{Fei2021}
When we select the remaining arbitrary function $\theta_{M+1}(z)$ as a constant function, oscillations appear around the exact spectral function.
As proposed in the previous study~\cite{Fei2021}, to remove oscillations, we expand $\theta_{M+1}(z)$ on the Hardy basis.
The coefficients are optimized by minimizing the cost function $|S-\int^\infty_{-\infty} \, d\omega \, \tilde{\rho}_A(\omega)|^2 + \lambda \int^\infty_{-\infty}\tilde{\rho}''_A(\omega)$, i.e., the Hardy optimization.
The first term enforces the sum rule of $\tilde{\rho}(\omega)$ and the second term suppresses oscillations.
In the Hardy optimization, we adopted an automatic differentiation approach with the \texttt{Zygote.jl} package~\cite{Innes2018} and minimized the cost function using the \texttt{Optim.jl} package~\cite{Mogensen2018} with $\lambda=10^{-4}$. 

In Fig.~\ref{fig:merge} (b), we show the resultant auxiliary fermionic spectral function, which behaves like a Hubbard gap structure.
In the continued data without Hardy optimization, oscillations are observed. 
However, with the Hardy optimization, the exact smooth spectral function can be reconstructed accurately.
By multiplying the fermionic spectral function with the hyperbolic tangent $\tanh(\beta\omega/2)$, we obtain the bosonic spectral function shown in Fig. ~\ref{fig:merge} (c).
A good agreement can be observed between the exact spectral function and the continued value.
This demonstrates that the Nevanlinna AC method can be applied to bosons with the hyperbolic tangent.

In conclusion, we demonstrated a method that converts bosonic and fermionic AC using a hyperbolic tangent.
We applied the Nevanlinna AC method to an auxiliary fermionic model and successfully reconstructed the original bosonic spectral function.
In realistic cases, sometimes we encounter numerical instabilities in Hardy optimization.
The development of robust optimization algorithms is left for future research.

\begin{acknowledgment}
The authors thank E. Gull, T. Koretsune, S. Namerikawa, F. Kakizawa, and Y. Yanase for their fruitful discussions.
K.N. was supported by JSPS KAKENHI (grant no. JP21J23007), and Research Grants (2022) of the WISE Program, MEXT.
H.S. was supported by JSPS KAKENHI Grants No. 18H01158, 21H01041, and 21H01003 and JST PRESTO Grant No. JPMJPR2012, Japan.
We used computational code based on \texttt{SprseIR.jl}. ~\cite{Wallerberger2023}
\end{acknowledgment}

\bibliographystyle{jpsj}
\bibliography{paper}

\end{document}